\documentclass[natbib]{svjour3}             
\smartqed  
\usepackage{graphicx}
 \usepackage{aps-bibstyle}  
\begin{document}

\title{Can surface flux transport account
       for the weak polar field in cycle 23?}


\author{Jie Jiang         \and
        Robert H. Cameron \and Dieter Schmitt \and Manfred Sch\"{u}ssler
}


\institute{Jie Jiang \at Key Laboratory of Solar Activity, National
Astronomical Observatories, Chinese Academy of Sciences, Beijing
100012, China \\ Max-Planck-Institut f\"ur Sonnensystemforschung,
37191 Katlenburg-Lindau, Germany \\ \email{jiejiang@nao.cas.cn} \and
Robert H. Cameron \and Dieter Schmitt \and Manfred Sch\"{u}ssler \at
Max-Planck-Institut f\"ur Sonnensystemforschung, 37191 Katlenburg-Lindau, Germany \\
\email{cameron@mps.mpg.de; schmitt@mps.mpg.de; schuessler@mps.mpg.de} 
}

\date{Received: date / Accepted: date}

\maketitle

\begin{abstract}
To reproduce the weak magnetic field on the polar caps of the Sun
observed during the declining phase of cycle 23 poses a challenge to
surface flux transport models since this cycle has not been
particularly weak. We use a well-calibrated model to evaluate the
parameter changes required to obtain simulated polar fields and open
flux that are consistent with the observations. We find that the low
polar field of cycle 23 could be reproduced by
   an increase of the meridional flow by 55\% in the last cycle.
   Alternatively, a decrease of the mean tilt angle of sunspot groups by
   28\% would also lead to a similarly low polar field, but cause a
   delay of the polar field reversals by 1.5 years in comparison to the
   observations.
\end{abstract}

\section{Introduction}
\label{intro} The current solar minimum shows some peculiar
features. There were 265 and 261 spotless days in the years 2008 and
2009, respectively. Only the years 1878, 1901 and 1913 showed more
spotless days in the daily records since 1849%
\footnote{http://users.telenet.be/j.janssens/Spotless/Spotless.html\#Year}.
The polar field is about 40\% weaker than during the previous three
minima for which routine magnetograph measurements are available%
\footnote{http://wso.stanford.edu/Polar.html}. Polar coronal-hole
areas shrank by about 20\% \citep{wang09} and the heliospheric open
flux reached the lowest values since systematic measurements started
in 1967 \textbf{\citep{wang09}}. Observations from Ulysses showed
that the high-latitude solar wind is 17\% less dense and cooler than
during the previous solar minimum \citep{McComas08}.  The unusual
properties of the polar corona, the open flux, and the solar wind
are probably all closely related with the weakness of polar field.

This situation poses an interesting challenge to surface flux
transport models (SFTM), which consider the passive transport and
evolution of the radial component of the magnetic field on the solar
surface under the effect of differential rotation, meridional flow
and diffusion \citep{Wang89, Ballegooijen98, Schrijver01, Mackay02,
Baumann04, Baumann06}. Such models have been shown to be consistent
with the observed evolution of the magnetic field distribution at
the surface, including the reversals of the polar fields. The models
take their input from the observed flux emergence (e.g., in terms of
sunspot data or magnetograms), so that their results are closely
tied to the activity level of the solar cycles. Since cycle 23 was
not particularly weak in comparison to its predecessors, the polar
fields generated by the flux transport models tend to be much
stronger than observed. In this paper, we consider which changes of
the parameters or input data would be required for cycle 23 in order
to bring the models into agreement with the actual data.

\section{Solar surface flux transport model}
\label{sec:1} 
\begin{figure}
\includegraphics[width=0.4\textwidth]{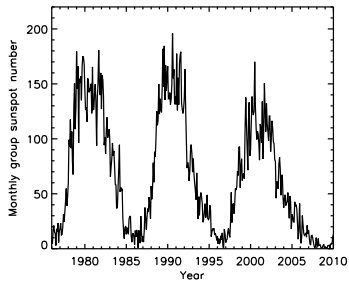}
\caption{Monthly international sunspot number since 1976.}
\label{fig:sn}       
\end{figure}

In a preceding study \citep{Cameron10}, we have
carried out flux transport simulations including the observed
cycle-to-cycle variation of sunspot group tilt angles to model the
solar surface field and open flux from 1913 to 1986, based upon the
RGO and SOON sunspot area data sets
\footnote{http://solarscience.msfc.nasa.gov/greenwch.shtml}. We
obtained a reasonable agreement of the time evolution of the open
flux and of the reversal times of the polar fields with empirical
results. Unfortunately, this model cannot easily be extended to
cover the last cycles because systematic observations of tilt angles
are not available and the definition of sunspot groups is
inconsistent between the RGO and SOON data sets. We have therefore
used the monthly international sunspot number since 1976 as the
basic input quantity describing the emergence of new flux (cf.
Fig.~\ref{fig:sn}). This method has been validated by comparison
with \citet{Cameron10} from 1913 to 1986 \citep[for details,
see][]{Jiang11b}.

\begin{figure}
\includegraphics[width=0.4\textwidth,angle=90]{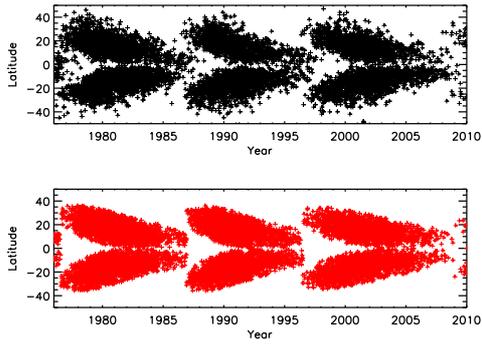}
\caption{Time-latitude distribution (butterfly diagram) of the
bipolar magnetic regions used as input for the SFTM (lower panel) in
comparison to the actual sunspot butterfly diagram from the SOON
data (upper panel).}
\label{fig:butterly}       
\end{figure}

We take the number of new bipolar magnetic regions (sunspot groups)
during a given month to be proportional to the corresponding sunspot
number.  The SFTM then requires the prescription of latitude,
longitude, area, and tilt angle for each new bipolar magnetic
region. This is done on the basis of empirical relations with cycle
strength together with using random numbers \citep[for details,
see][]{Jiang11a}. As an example, Figure~\ref{fig:butterly} shows the
time-latitude distribution of emerging bipolar regions as model
input for the three cycles considered here in comparison with the
actually observed butterfly diagram.

The area of bipolar regions is chosen randomly according to the
observed size distribution function \citep{Jiang11a}, with a higher
probability for large groups during cycle maximum phases
\citep{Hathaway10}. Since preferred longitudes of flux emergence
(activity nests) strongly affect the equatorial dipole component and
thus the open flux during solar maxima, we draw the emergence
longitudes from a combination of random longitudes with a set of
longitudes clustering on two values 180 degrees apart \citep[for
details, see][]{Jiang11a,Jiang11b}.
On the basis of Mount Wilson and Kodaikanal sunspot data,
\citet{Dasi-Espuig10} found that the average tilt angle of sunspot
groups is negatively correlated with the strength of the cycle. The
cycle-dependent factor $T_n$ in the fit relation
$\alpha_n(\lambda)=T_n\sqrt{\lambda}$ ($\alpha_n$: average tilt
angle for cycle $n$; $\lambda$: latitude) is determined by using
this anti-correlation \citep{Cameron10}. The values of $T_n$ for
$n=21$, 22, and 23 are 1.21, 1.21, and 1.32 \citep{Jiang11b}.

\begin{figure}
\includegraphics[width=0.4\textwidth]{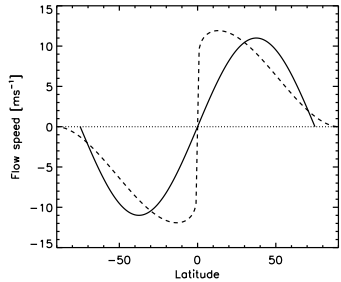}
\caption{Profile of meridional flow adopted in our reference model
(solid curve) and the profile used by \citet[][dashed curve]{wang09}.}
\label{fig:mf}       
\end{figure}

The profile for the differential rotation used in our SFTM
simulations is
$\Omega(\lambda)=13.38-2.30\sin^{2}\lambda-1.62\sin^{4}\lambda$
\citep[in degrees per day;][]{Snodgrass83}. The profile for the
meridional flow is \citep[cf.][]{Ballegooijen98}
\begin{equation}
\upsilon(\lambda)=\left\{
  \begin{array}{l l}
     11\sin(2.4\lambda) ~\mathrm{ms^{-1}} & \mathrm{where} |\lambda| \leq 75^{\circ}\\
     0 & \textrm{otherwise},
  \end{array}
  \right.
\end{equation}
which is shown by the solid line in Figure \ref{fig:mf}. The
supergranular turbulent diffusivity is chosen as
$250\,\mathrm{km^2s^{-1}}$ \citep[following][Table
6.2]{Schrijver:Zwaan:2000}.

The initial magnetic field distribution at the start of the
simulation is given by Eq. 10 of \citet{Cameron10} (see also van
Ballegooijen et al. 1998). The amplitude of the initial field has
been set such that the line-of-sight field averaged over a
35$^\circ$ wide polar cap agrees with the corresponding WSO data for
1976 (see Fig.~5). A decay term due to radial diffusion of the
magnetic field \citep{Baumann06} has not been considered here since
the observed anti-correlation of sunspot group tilt angle and cycle
strength \citep{Dasi-Espuig10} removes the necessity for this term
in short-term studies \citep{Cameron10}. These choices of the model
parameters and the flux input constitute our `reference model'.

\section{Results of the reference model}
\begin{figure}
\includegraphics[width=0.4\textwidth,angle=90]{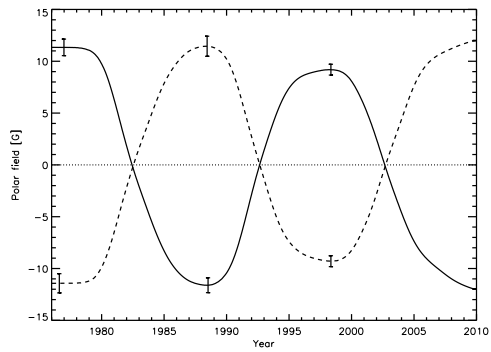}
\caption{Simulated evolution of radial component of solar north
(full line) and south (dashed line) polar field, averaged over the
$15^{\circ}$-wide polar caps. The error bars denote the standard
deviation for 20 runs with different sets of random numbers.}
\label{fig:pf_br}       
\end{figure}

\begin{figure}
\includegraphics[width=0.4\textwidth,angle=90]{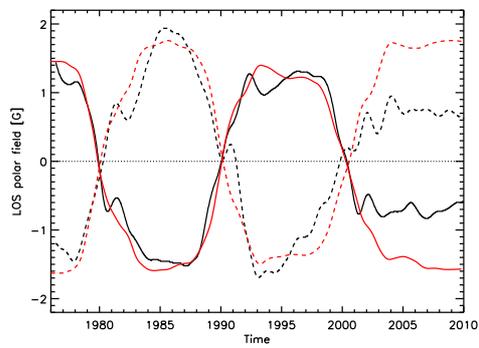}
\caption{Comparison between the simulated (red curves) and observed
(WSO data, multiplied by a factor 1.3, black curves) line-of-sight
component of solar polar field, averaged over $35^{\circ}$-wide
polar caps. Full lines correspond to the north pole, dashed lines to
the south pole.}
\label{fig:pf_los}       
\end{figure}

Figure \ref{fig:pf_br} shows the time evolution of the simulated
radial component of the polar fields, averaged over
$15^{\circ}$-wide polar caps. Since the location, size and emergence
time of the bipolar magnetic regions involve random numbers, the
results from different sets of random numbers vary somewhat. The
error bars in Figure \ref{fig:pf_br} denote the standard deviation
for 20 runs with different random numbers. In order to compare with
the line-of-sight polar field above $55^{\circ}$ latitude observed
at Wilcox Solar Observatory (WSO), we use the same definition of the
polar field as WSO. The result is shown in Figure \ref{fig:pf_los}.
The WSO polar field is multiplied by a constant factor 1.3 to
correct for magnetograph saturation \citep{Svalgaard78}. There is
good agreement between the simulation and the measurement for both
the amplitude and reversal times of the polar field between 1976 and
2002. However, during the end of cycle 23, the simulation gives
polar fields that are too strong by a factor of about 2.

\begin{figure}
\includegraphics[width=0.4\textwidth,angle=90]{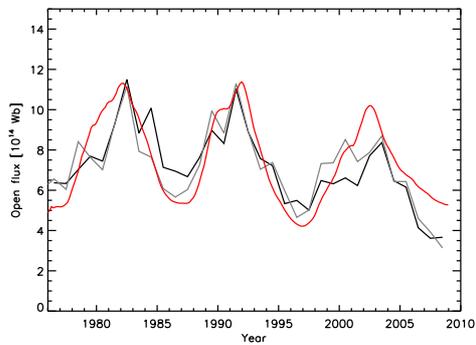}
\caption{Evolution of heliospheric open flux since 1976. Shown is
the result of the reference SFTM (red curve) in comparison to the
OMNI spacecraft data \citep[with kinematic correction,][black
curve]{lock09b} and values derived from geomagnetic variations
\citep[with kinematic correction,][grey curve]{lock09a}.}
\label{fig:of}       
\end{figure}

In addition to the polar field, we also use the heliospheric open
flux to compare observation and SFTM simulation.  To this end, we
extrapolate the simulated photospheric field out to the source
surface using the current sheet source surface (CSSS) model
\citep{Zhao95a, Zhao95b, Jiang10} with the model parameters given by
\citet{Cameron10}. Figure \ref{fig:of} shows the comparison between
the simulation results and the OMNI spacecraft data \citep[with a
kinematic correction, cf.][]{lock09b}.

The simulated open flux is consistent with the measurements in both
maximum and minimum phases before 2005. During the minimum of cycle 23,
however, the simulated result is much higher than the measurement. This
is directly related to the too strong polar field since the open flux
during activity minima is dominated by the low-order axial multipoles.

\section{Possible cause of the weak polar field in cycle 23}
What could have been different in cycle 23?  Which changes in the
SFTM parameters and/or input properties could possibly lead to the
low polar field? The turbulent diffusivity models the random walk of
magnetic elements due to supergranular motions. Since these motions
are observed to only weakly vary over the solar cycle \citep[for a
review see][ section 4.6.5]{Rieutord10}, we have assumed that the
diffusivity is constant in time. Likewise, the time-latitude
distribution of emerging flux during cycle 23 is very similar to
that of the preceding cycles (cf. Fig.~\ref{fig:butterly}). The same
is true for the mean sunspot group areas as derived from the SOON
data .

We consider three possibilities for potential parameter variations.
a) sunspot number: \citet{Wilson05} found that the international
sunspot number has been systematically overestimated since 1981.
Lower values of the sunspot number would entail less flux emergence
in our model and thus a lower polar field. b) tilt angle: although
\citet{Schrijver08} found no indication for a systematic change in
the mean tilt angle by analyzing selected MDI magnetograms, one has
to keep in mind that the determined tilt angles are sensitive to the
location and evolution phase \citep{yang09} of the sunspot group
considered, to the observer's definition, and to the instrument.
Furthermore, the tilt angles show a very large scatter, possibly due
to the convective buffeting of rising flux tubes in the upper layers
of the convection zone \citep{Longcope02}. c) meridional flow:
significant variability of the meridional flow has been detected
during cycle 23 \citep{Chou01, Gizon04, Hathaway:Rightmire:2010,
Gonzalez10}.  We considered variations of the meridional flow, of
the mean sunspot group tilt angle, and of sunspot number during
cycle 23 as potential causes of the low polar field during that
cycle and used our SFTM to determine which changes of these
properties in cycle 23 would be required to reproduce the
observations.

\begin{figure}
\includegraphics[width=0.4\textwidth,angle=90]{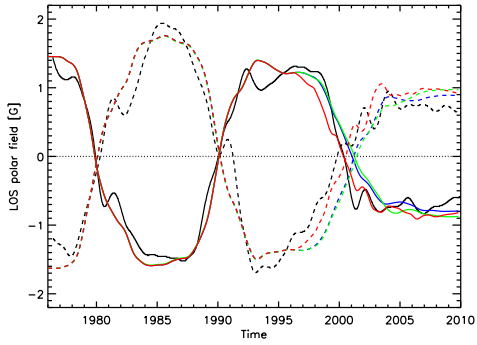}
\caption{Evolution of line-of-sight polar field with modified
parameters during cycle 23: 40\% decrease of sunspot number (green
curves), 28\% decrease of the sunspot group tilt angle (blue curves,
which nearly lie upon the green curves), 55\% increase of meridional
flow (red curves). The three simulated cases are identical before
the start of cycle 23. The WSO observations are given by the black
curves. In all cases, full lines correspond to the north pole,
dashed lines to the south pole.}
\label{fig:pfs}       
\end{figure}

\begin{figure}
\includegraphics[width=0.4\textwidth,angle=90]{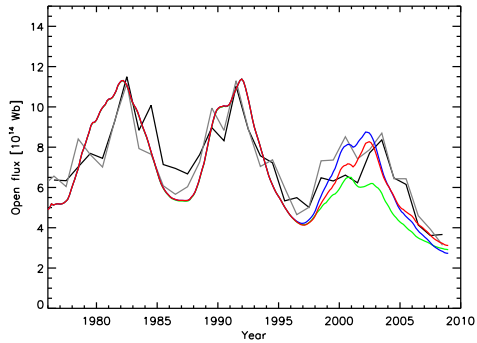}
\caption{Evolution of the heliospheric open flux with modified
parameters during cycle 23. The colored lines refer to the same
cases as in Figure~\ref{fig:pfs}. The black and grey lines give the
OMNI data and values derived from geomagnetic variations,
respectively. }
\label{fig:ofs}       
\end{figure}

Figures~\ref{fig:pfs} and \ref{fig:ofs} show the evolution of the
polar field and the heliospheric open flux, respectively, that
result from such changes (varying one parameter at a time): a) a
40\% reduction of the sunspot number (green curves), b) a 28\%
decrease of mean tilt angle (blue curves), or c) a 55\% increase of
the meridional flow (red curves), always with respect to the values
for cycle 23 in the reference model. In case a), the open flux
during the maximum phase of cycle 23 is much weaker than the
observed value since the decreased amount of flux emergence leads to
a strongly reduced equatorial dipole moment during the maximum
phase. This excludes an overestimate of the sunspot number as sole
cause of the low polar field. For case b), the reduced tilt angles
lead to a delay of the polar field reversal by about 1.5 yrs, which
is in disagreement with the observations. Smaller tilt angles
correspond to less meridional magnetic flux of the bipolar magnetic
regions, so that it takes longer (more bipolar regions) to reverse
the polar field of the preceding cycle.

For the case c) with increased meridional flow, both polar field
reversal time and amplitude match the observation (cf.
Figure~\ref{fig:pfs}). A stronger meridional flow, especially an
increased velocity gradient near the equator, effectively separates
the two hemispheres, so that the transport of net flux to the poles
is reduced. Although the contribution of each magnetic region to the
polar field is smaller, the transport velocity is higher, thus
minimizing the effect on the polar field reversal time. The
increased meridional flow also brings the simulated heliospheric
open flux into reasonable agreement with the observation during both
the maximum and the minimum phases of cycle 23 (cf.
Figure~\ref{fig:ofs}). This result is consistent with the SFTM
simulations of \citet{wang09} and \citet{Schrijver08}.
\citet{wang09} found that in their model already a 15\% increase of
the meridional flow, much smaller than our value of 55\%, is
sufficient to reproduce the observed weak polar field. This is
probably due to the extremely steep velocity gradient near the
equator in the meridional flow profile adopted by these authors (cf.
Figure~\ref{fig:mf}). However, such a profile does not seem to be
very realistic in the light of helioseismic measurements
\citep[e.g.,][]{Gizon05}.

\section{Conclusion}

Our simulations indicate that quite substantial changes in the SFTM
   parameters are required to obtain the observed weak polar field
   during cycle 23, at least if the effect is ascribed to a change in only
   one parameter, such as a 28\% decrease of the mean sunspot group tilt
   angle or a 55\% increase of the meridional flow (from 11~$\mathrm{m
   s^{-1}}$ to 17~$\mathrm{m s^{-1}}$). Note that a combination of
   changes in more than one parameter would of course require smaller
   variations to bring the SFTM simulations into agreement with the
   observations. For instance, the results of
   \citet{Hathaway10} indicate that the meridional flow
   speed during the decay phase of cycle 23 was nearly 20\% higher than
   during the corresponding phase in cycle 22. Furthermore, the
   meridional flow profile used in the flux transport simulation
   probably underestimates the poleward velocities beyond 60$^\circ$
   latitude \citep{Hathaway:Rightmire:2011}. On the
   other hand, polar countercells of the meridional flow could
   redistribute flux from the polar cap to lower latitudes and thus also
   contribute to weaker polar fields \citep{Jiang09}. When
   additional empirical data (e.g., concerning the tilt angles) become
   available, it will be possible to further constrain the simulation
   parameters and hopefully achieve a better understanding of this
   rather unusual solar minimum.

\begin{acknowledgements}
JJ thanks the organizers of Cosmic Rays in the Heliosphere II for an
invitation to participate and the financial support. The National
Basic Research Program of China through grants no. 10703007 and
10733020 are acknowledged for the partial funding of JJ's research.
\end{acknowledgements}

\bibliographystyle{aps-nameyear}      
\bibliography{cycle23}   
\nocite{*}

\end{document}